# Topological security assessment of technological networks

Enzo Fioriti[1], Marta Chnnici[1] and Andrea Arbore[2]

[1]ENEA, CR Casaccia, Roma
[2]Aletea SpA, Roma

*Abstract*—**The spreading of dangerous malware or faults in inter-dependent networks of electronics devices has raised deep concern, because from the ICT networks infections may propagate to other Critical Infrastructures producing the well-known domino or cascading effect. Researchers are attempting to develop a high level analysis of malware propagation discarding software details, in order to generalize to the maximum extent the defensive strategies. For example, it has been suggested that the maximum eigenvalue of the network adjacency matrix could act as a threshold for the malware's spreading. This leads naturally to use the spectral graph theory to identify the most critical and influential nodes in technological networks. Many well-known graph parameters have been studied in the past years to accomplish the task. Here we test our AV11 algorithm showing that outperforms degree, closeness, betweenness centrality and the dynamical importance.**

## I. INTRODUCTION

The malicious software (mal-ware) is a program code designed to produce undesired effects on a computer. Once malware was specialized. Today, the trend is reversed toward an unification of these different dangerous codes and towards a very high technical level. An issue is the capability to influence not only the ICT network but also various Critical Infrastructures dependent from ICT [4], [5], [6], [7], [8], [9], [11], [13]. In order to obtain such a result, there are basically two strategies: the targeted intrusion and the cooperative search. The first foresees a direct conventional approach to the actual target, while the second one demands a distributed control system, a complex communication scheme and a consensus-like decision making process. As a side effect of the cooperative search, the malware (or the fault, it is the same in our approach) will spread in the network like a disease (the "epidemic" spreading). Actually, any kind of worm follows the epidemic spreading, but a standard worm will attempt to invade the maximum number of machines as quickly as possible, instead a sophisticated malware adopting a cooperative search strategy or even a simpler network aware strategy, will infect (relatively) few machines during a long period of time. In any case, both seem to propagate following the epidemic spreading model, at least during the initial phase of the attack. Thus, understanding this model may help in countering the spreading at the very beginning of it, when the cost of the defence is more affordable. Important results on the threshold to the spreading are those of Pastor-Satorras and Vespignani [3] for the scale free graphs, and subsequently by Wang et al. [1] and Chakrabarti et al. [2] for a generic graph. "Generic graph" means no assumption is made on the graph (scale-free, random, small-word, degree distribution, etc) or on the modeling Susceptible-Infected-Susceptible, Susceptible-Infected-Refractory, Infected-Refractory, etc (SIS, SIR, IR).

The threshold is related to two parameters, namely the infection rate $\beta$ (average number of machines that can be infected per time unit by an already infected machine) and the cure rate $\delta$ (average number of machines that can be restored per time unit). Above the threshold the malware will propagate, otherwise it will end quickly.

We will use this claim to identify the most influential subset of nodes with respect to the maximum eigenvalue.

## II. DETERMINING THE THRESHOLD

We will sketch the calculation of the threshold using the Peng's framework, as a theoretical support of our immunization strategy. Peng et al. [7] have provided an analytical treatment when $\beta$ and $\delta$ vary but have tested their claim only on a random Erdos-Renyi artificial graph. Here we outline briefly the formalism to derive the stability conditions for the dynamic system of the spreading, (for details see [7]). The homogeneous models [1],[2],[3] assume that every machine has equal contact to others in the population, thus the infection and cure rates are constant; instead, in this paper we consider a different infection rate for each link $\beta_{ij}$ and a different cure rate for each node $\delta_i$ of the (directed or undirected) graph G representing the Italian AS network. $\beta_{ij}$ and $\delta_i$ are extracted from a uniform distribution. The first step [7] is the modified adjacency matrix **M** obtained from the standard adjacency matrix **A** of the undirected graph G, whose entries $m_{ij}$ are modified according to:

$$m_{ij} = \beta_{ij} \quad \text{if } i \neq j, \quad 0 \leq \beta_{ij} \leq 1$$
$$m_{ij} = 1 - \delta_i \quad \text{if } i = j, \quad 0 \leq \delta_i \leq 1$$

Note that we allow G to be directed, i.e. $\beta_{ij} \neq \beta_{ji}$ with no loss of generality. The difference system representing the infection dynamics on G is:

$$p_i(t) = 1 - \prod_k (1 - m_{i,k} \cdot p_k(t-1)), \quad i, k=1 \ldots N \quad (1)$$

where $p_i(t)$ is the probability that the node $i$ at the discrete time $t$ is infected from the node $k$, $N$ is the number of nodes. Note that the $p_i(t-1)$ should be mutually independent; if this is not the case, the threshold value cannot be calculated exactly within the Peng's framework [7]. Now, since:

$$1 - \prod_k (1 - m_{i,k} \cdot p_k(t-1)) < \prod_k (m_{i,k} \cdot p_k(t))$$

the system (2.1) converges to zero iff the difference system (2.2)

$$p_i(t) = \prod_k (m_{i,k} \cdot p_k(t-1)), \quad i, k=1 \ldots N \quad (2)$$

converges to zero. In compact notation the (2.2) is:

$$\mathbf{P}(t) = \mathbf{M} \cdot \mathbf{P}(t-1) \quad (3)$$

The system (3) is stable if the largest eigenvalue of $\mathbf{M}$ is [9]:

$$\lambda_\mathbf{M} < 1$$

therefore:

$$\lim \mathbf{P}(t) = 0 \quad \text{when } t \to \infty$$

and

$$p_i(t) \to 0, \quad i = 1 \ldots N$$

the epidemic spreading disappears. Since $\mathbf{M}$ is non negative, its largest eigenvalue is a real number and the analytical threshold can be set to:

$$\lambda^{thr}_\mathbf{M} = 1 \quad (4)$$

Anyway, in this paper we do not use this threshold due to the *unrealistic* [7] independence assumptions for (1).

Note that (4) says nothing about the actual spreading *above* the threshold: it states only a stop *below* the threshold. The real significance of (4) is that *the spreading depends on graph topology*: immunizing nodes/links affects strongly the diffusion of malware. It is also worth noting that the threshold may be considered as a phase transition, whose mathematical treatment is difficult [15].

### III. THE IMMUNIZATION ALGORITHMS

Now we introduce some of the most popular procedures to determinate a number of the most influential nodes [8] (called the "budget"). Once these nodes have been selected, we "immunize" them, meaning the cure rate becomes high. Then we run the simulations and verify if the infection has been stopped or mitigated. All the methods (except the k-core, because it depends directly from the degree) have been tested to evaluate their immunization capability. Actually, many more methodologies exist, but the following are probably the more representatives.

**Most infected**: according to this simple procedure, the most influential nodes are those who get infected very often in the simulations.

**Degree**: simply the number of links form/to a node. Intuitively a high degree node ("hub") may be influential.

**Closeness**: inverse measure of centrality associated with a node. The sum of the shortest path lengths between a given node and all other nodes in the graph. Vertices that tend to have short geodesic distances to other vertices in the graph have higher closeness.

**Betweenness**: total number of shortest paths between every possible pair of nodes that pass through the given node. Vertices that occur on many shortest paths between other vertices have higher betweenness.

**Dynamical Importance**: variation of the max eigenvalue after a node has been removed. Indicates how much the node is influential with respect to the others. It is a spectral method.

**Estrada Index**: SC of a vertex is the "sum" of closed walks of different lengths in the network starting and ending at vertex. It characterizes nodes according to their participation in subgraphs of the network, giving higher weights to the smaller subgraphs that can be involved in network motifs. It is a spectral method.

**K-core**: every subgraph has a vertex of degree at most $k$: That is, some vertex in the subgraph touches $k$ or fewer of the subgraph's edges. Nodes are ranked accordingly.

**AV11**: For a given graph G, we want to identify simultaneously the $k$ best nodes (the "budget") to immunize or remove, to make the remaining nodes more robust to the attack.

Of course, following the spectral paradigm one could remove a set of $k$ nodes and find the decrease of the eigenvalue, but this brute-force strategy is impossible to use, even for small graphs, because of the huge number of combinations. The problem is NP-complete [14], thus we resort to our suboptimal algorithm AV11 to reduce complexity and calculation time as its estimated complexity is $O(kn^3 * \lg(k))$. On the other hand, the *simple strategy* (calculating the eigenvalue for a single node at a time, rank the results and take the first $k$ nodes) does not guarantee good performances, and above all, is easily understood by the attacker. In fact the attacker running the same simple algorithm would get exactly the same information. The AV11 pseudocode (see the Appendix for a justification and [8] for details) is:

1. Calculate eigenvalues of the adjacency matrix A;
2. Initialize: **S** to empty; $\mathbf{Z} = \mathbf{I}_n$; $\mathbf{D} = (1+|\lambda_{m1}|)\mathbf{I}_n$;
3. Node=0. For $i$= 0 to $k$ do
4. $\mathbf{P} = (\mathbf{Z}*\mathbf{A}*\mathbf{Z}+\mathbf{D})^p$;
5. Node= max( diag(**P**) );
6. Add node to **S**;
7. Set **Z**[node, node]= **0**;
8. end for ;
9. return **S**.

## IV. EXPERIMENTS

In Table 1 and Figure 1 are shown the effects of various immunization strategies. Starting from the network topology of 760 nodes of a LAN (Local Area Network), the algorithm select a number of nodes to be immunized (the "budget"). Clearly the budget should be as small as possible: in this case is below 4%. The infection developed in the center of the LAN (top image in Figure 1), then all the nodes were affected. In the middle image are shown the nodes selected by AV11 to be immunized; in the last image the red points are the computers remained infected after the AV11 immunization.

Other experiments have been conducted on a High Voltage electricity distribution network (Table 2 and Figure 2), following the same criteria as above. Here the budget is the 13% of 208 nodes.

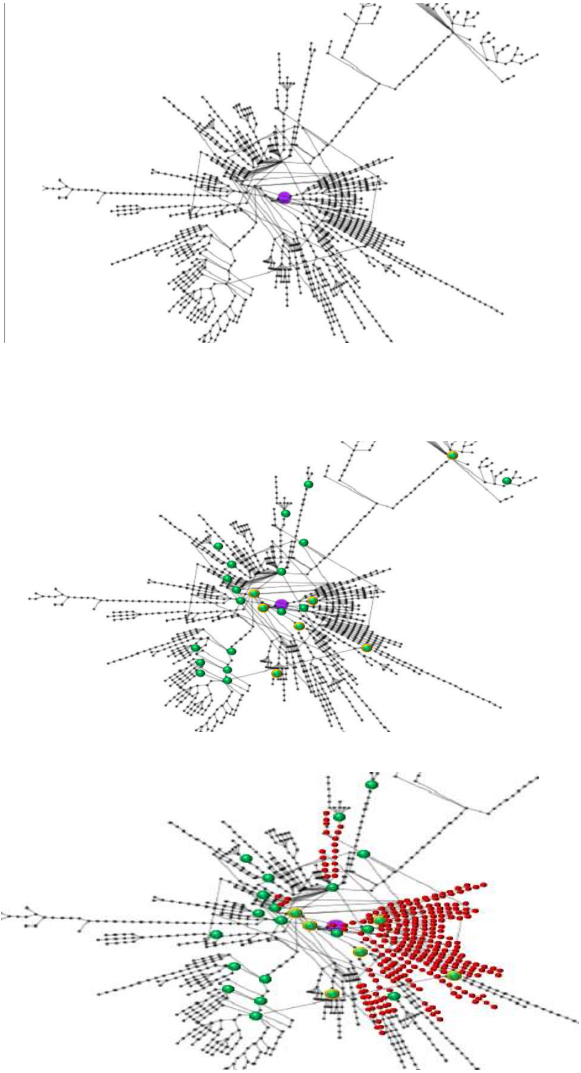

Fig. 1. The malware diffusion in a LAN and the effect of the AV11 immunization (red points are the remaining infected nodes). The overall number of nodes in the LAN is 760. The budget is below 4% of nodes. The blue large point in the centre of the upper imagine is the node who started the infection. The green points are the nodes selected by AV11 as immunization budget (the green-yellow points are also hubs).

TABLE I
The results of the immunization procedures in the LAN

| Algorithm | Nr. still infected nodes, (percentage) |
|---|---|
| AV11 | 176 (23%) |
| ESTRADA INDEX | 258 (34%) |
| CLOSENESS | 263 (35%) |
| DEGREE | 377 (36.5%) |
| DYNAMICAL IMPORTANCE | 394 (52%) |
| BETWEENNESS | 414 (54.5%) |
| MOST INFECTED | 661 (87%) |

TABLE 2
The results of the immunization procedures in the HV distribution

| Algorithm | Nr. still infected nodes, (percentage) |
|---|---|
| AV11 | 5 (2.4%) |
| DEGREE | 10 (4.8%) |
| BETWEENNESS | 22 (10.6%) |
| ESTRADA INDEX | 36 (17.4%) |
| CLOSENESS: | 97 (47%) |
| MOST INFECTED: | 100 (48.4%) |
| DYNAMICAL IMPORTANCE | 146 (70.5%) |

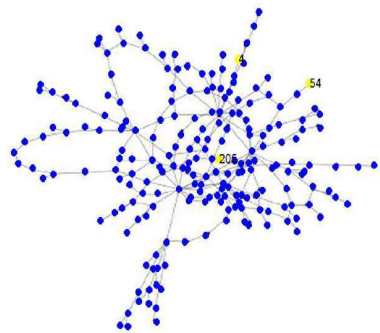

Fig. 2. An high voltage distribution grid. The simulation considers a cascading failure propagating into the grid. The yellow points initiate the cascading process. Note how the grid is well structured in order to guarantee communications and energy transmission, but at the same time, the connectivity facilitates the cascading effects.

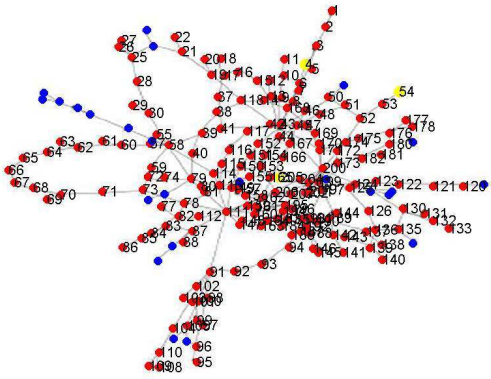

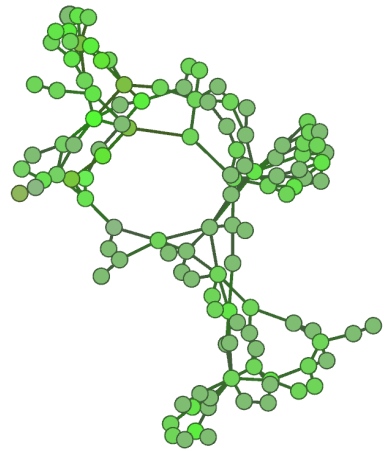

Fig. 4. The IEEE 118 BUS distribution grid. Before the immunization, almost all nodes were affected by a cascading failure. The budget is the 16% of 118 nodes.

Fig. 3. An high voltage distribution grid. Before the immunization, almost all the nodes were affected by a cascading failure (first image, red points). After the AV11 immunization, only 5 nodes remained affected by the cascading failure (second image). The budget was 13% of 208 nodes.

Finally, the IEEE 118 Bus is analyzed (see Table 3 and Figure 4). The budget of nodes to be immunized is 16% of 118 nodes.

TABLE 3
The results of the immunization procedures in the IEEE 118 Bus

| Algorithm | Nr. still infected nodes, (percentage) |
|---|---|
| AV11 | 9 (7.6%) |
| DEGREE | 16 (13.6%) |
| BETWEENNESS | 52 (44%) |
| ESTRADA INDEX | 60 (50.8%) |
| CLOSENESS | 70 (59%) |
| MOST INFECTED | 85 (72%) |
| DYNAMICAL IMPORTANCE | 88 (74%) |

## V. CONCLUSION

The AV11 algorithm performs better than the others parameters in different topological scenarios. Note that the second best is not always the degree centrality, as an intuitive approach could conclude. We think the same holds for all the degree-like parameters as the K-core.

Even more important, note the very poor performance of the most infected strategy. According to this procedure, the influential nodes, i.e. those nodes that provide a major support to the epidemic spreading or the cascade effect, are those who get infected often during simulations. Our results demonstrate how deceiving this idea is.

Today the infrastructure protection is a major issue in the most important research programs in the world. In particular, the ICT infrastructure undergoes devastating attacks generated by malware and propagating to other Critical Infrastructures following a domino effect pattern. Software attack codes are becoming extremely sophisticated, difficult to detect or foresee and can adopt a network aware strategy or even an advanced cooperative target search.

Although more simulations should be repeated on other ICT networks, these tests suggest clearly that the spreading in the technological networks is governed by the topology, although our results may be transferred to different type of graph as the contact graph, but probably not to the social networks, who are governed by more complex interactions.

Therefore, modelling the malware spread or the fault cascading on graphs may result in a general purpose passive defence scheme, effective against a broad range of threats.

## APPENDIX

Because an adjacency matrix **A** is symmetric, its eigenvalues are real. Let $\lambda_1 \geq \lambda_2 \geq ... \geq \lambda_j \geq \lambda_{j+1} \geq ... \geq \lambda_{n-1} \geq \lambda_n$ be the $n$ eigenvalues of **A**. We apply the Separation Theorem for the Hermitian operators on an $n$-dimensional space. From this classical theorem, follows that if $\mathbf{A}_r$ is a principal submatrix of **A** of order $r$ with eigenvalues $\alpha_1 \geq$

$α_2 ≥...≥ α_{r-1} ≥ α_r$ then $λ_j ≥ α_j ≥ λ_{n-r+j}$. Therefore, removing $k$ nodes from the network we get a $n-k$ principal sub-matrices, whose eigenvalues are localized in $λ_j ≥ α_j ≥ λ_{n-(n-k)+j} = λ_{k+j}$. For the largest eigenvalue of $\mathbf{A}_r$, we obtain $λ_1 ≥ α_1 ≥ λ_{k+1}$. Thus $λ_{k+1}$ is the absolute minimum largest eigenvalue we can obtain deleting $k$ nodes from the network. It is well known that the largest eigenvalue may be overestimated with the power method. The basics of the method are the following: at iteration $h$, the left and right products $\mathbf{Z} = f(h)$ reset $h$ rows and $h$ colomns of $\mathbf{A}$ ($h$ nodes removed). The result $\mathbf{A}_{(h)}$ is the $n×n$ adjacency matrix of the graph with $h$ nodes isolated (or immunized). The positive diagonal shift by $\mathbf{D} =(1+|λ_n|)\mathbf{I}_n$ guarantees that every eigenvalue $β_j(h)= d + λ_{j(h)}$ is positive, where $λ_{j(h)}$ is the $j$-th eigenvalue of $\mathbf{A}(h)$. Now, the trace of the power is:

$$(\mathbf{Z}·\mathbf{A}·\mathbf{Z}+\mathbf{D})^p = (\mathbf{A}_{(h)} + d·\mathbf{I}_n)^p = (b_{11(h)} … b_{1n(h)} … … … b_{1n(h)} … b_{nn(h)})$$

and it is known to satisfy:

$$\text{Trace} = (\mathbf{A}_{(h)} + d·\mathbf{I}_n)^p > \sum_i b_{ii}(h) ≥ \sum_j (d+ λ_{j(h)})^p$$

Therefore, from:

$$(d+λ_{1(h)})^p [1+ \sum_j (d+ λ_{j(h)} / d+ λ_{1(h)})^p] = \text{Trace}(\mathbf{A}(h) + d·\mathbf{I}_n)^p$$

follows that:

$$(d+ λ_{1(h)}) [1+ \sum_j (d+ λ_j(h) / d+ λ_{1(h)})^p]^{1/p} = (\text{Trace}(\mathbf{A}_{(h)} + d·\mathbf{I}_n))^{1/p}$$

Being $0 < (d+ λ_{j(h)}) / d+ λ_{1(h)}) < 1$

$j=2,3,...,n$

And when $p→∞$, we have:

$$d+ λ_{1(h)} ≈ (\text{Trace}(\mathbf{A}_{(h)} + d·\mathbf{I}_n)^p)^{1/p}$$

Here we are interested in the:

$$d + λ_{1(h)} ≤ (\text{Trace}(\mathbf{A}_{(h)}+ d·\mathbf{I}_n)^p)^{1/p} = (\sum_i b_{ii(h)})^{1/p}$$

In words, (A.4) states that reducing diagonal elements $b_{ii(h)}$ of $(\mathbf{A}_{(h)}+d·\mathbf{I}_n)^p$ forces the reduction of the largest eigenvalue $λ_{1(h)}$. Resetting the $i$-th row and the $i$-th colomn in $\mathbf{A}$, means to reset an eigenvalue $λ_{j(h)}$ and make $b_{ii(h)} = d^p$

Hence, at $h$ iteration there are $h$ diagonal elements such that $b_{ii(h)} = d^p$ and

$$d+ λ_{1(h)} ≤ (hd^p + \sum_j b_{ij\ ij\ (h)})^{1/p}$$

The positive diagonal shift $D =(1+|λ_n|)\mathbf{I}_n$ allows to use any positive integer p for the power. In fact, if p is an odd number and some $d+λ_{j(h)}$ are negative, then the ratios $d+λ_{j(h)} / d+λ_{1(h)}$ are not all positive; this invalid the inequality. In addition, for the Separation Theorem we have $d+α_j > 0$ for every principal sub-matrix of $\mathbf{A}$. Thus, we can use the same value during all the iterations: every shifted adjacency matrix will be a positive definite matrix. The computational complexity of AV11 is $O(kn^3\log(p))$ as the power of a matrix is computed by repeated multiplications. If the complexity for each product of two $n×n$ matrices is $O(n^3)$, then we have to do $O(n^3\log(p))$ operations for $k$ iterations.